\documentclass[conference]{IEEEtran}
\usepackage{amssymb,amsmath,amsthm}
\usepackage{graphicx}
\usepackage{subcaption}
\usepackage{subfig}
\usepackage{mdwmath}
\usepackage{mdwtab}
\usepackage{soul}
\usepackage{cite}
 \usepackage{multirow}
 \usepackage{color} 
  \usepackage[linesnumbered,ruled]{algorithm2e}
\usepackage{setspace}
\usepackage{esvect}
\usepackage{amsmath}

\begin{document}
%
\title{Balancing the Demands of Reliability and Security with Linear Network Coding in Optical Networks}
\author{\IEEEauthorblockN{Anna Engelmann and Admela Jukan}
\IEEEauthorblockA{Technische Universit\"at Carolo-Wilhelmina zu
Braunschweig, Germany\\
Email: \{a.engelmann, a.jukan\} @tu-bs.de}
}
\maketitle

\begin{abstract}
Recently, physical layer security in the optical layer has gained significant traction. Security treats in optical networks generally impact the reliability of optical transmission. Linear Network Coding (LNC) can protect from both the security treats in form of eavesdropping and faulty transmission due to jamming. LNC can mix original data to become incomprehensible for an attacker  and also extend original data by coding redundancy, thus protecting a data from errors injected via jamming attacks.  In this paper, we study the effectiveness of LNC to balance reliable transmission and security in optical networks. To this end, we combine the coding process with data flow parallelization of the source and propose and compare optimal and randomized path selection methods for parallel transmission. The study shows that a combination of data parallelization, LNC and randomization of path selection increases security and reliability of the transmission.  We analyze the so-called \emph{catastrophic security treat} of the network and show that in case of conventional transmission scheme and in absence of LNC, an attacker could eavesdrop or disrupt a whole secret data by accessing only one edge in a network.
\end{abstract}


%

\section{Introduction}
\par Recently, physical layer security in the optical layer has gained significant traction. Security treats in optical networks generally impact the reliability of optical transmission. In the case of fiber cut and tapping attacks, an attacker has physical access to network equipment and can disrupt the operation. The interchannel eavesdropping and jamming, on the other hand, can be lunched from a legitimately acquired optical channel \cite{Yuan:2014}. In addition, an attacker can exploit the crosstalk effects and launch both jamming and eavesdropping attacks. For a jamming attack, the attacker injects a high-power jamming signal ($5-10$ dB above other, legitimate signals) into the acquired lightpath inducing a severe interference effect for transmission channel on the fiber link, thus degrading the signal quality. At the same time, the attacker can collect the leakage signals from adjacent lightpaths and extracts the secret information by amplifying the collected signals \cite{Furdek:2014}.

\par A number of effective solutions against jamming and eavesdropping attacks propose encryption and optimization of lightpath disjointness; encryptions can effectively protect against eavesdropping, and path disjointness can make eavesdropping more difficult. One of the prospective methods used in networks other than optical includes linear network coding (LNC).  LNC can protect from both the jamming and eavesdropping. First, the source combines the original data with random information and designs a network code in such a way that only the receivers are able to decode the original packets. The nodes can only decode packets if they have received a sufficient number of linearly independent information vectors, which an eavesdropper might not be able to do. Second, LNC is also an erasure coding technique and thus can protect the source message also from random errors, and errors injected by the wiretapper via jamming attacks \cite{Cai:2011,Fragouli:2006}. LNC can provide the so-called \emph{r-secure coding,} whereby a wiretapper can eavesdrop any $r$ channels in the network, without gaining any knowledge about the source message. 
\begin{figure*}[!ht]
\includegraphics[width=2\columnwidth]{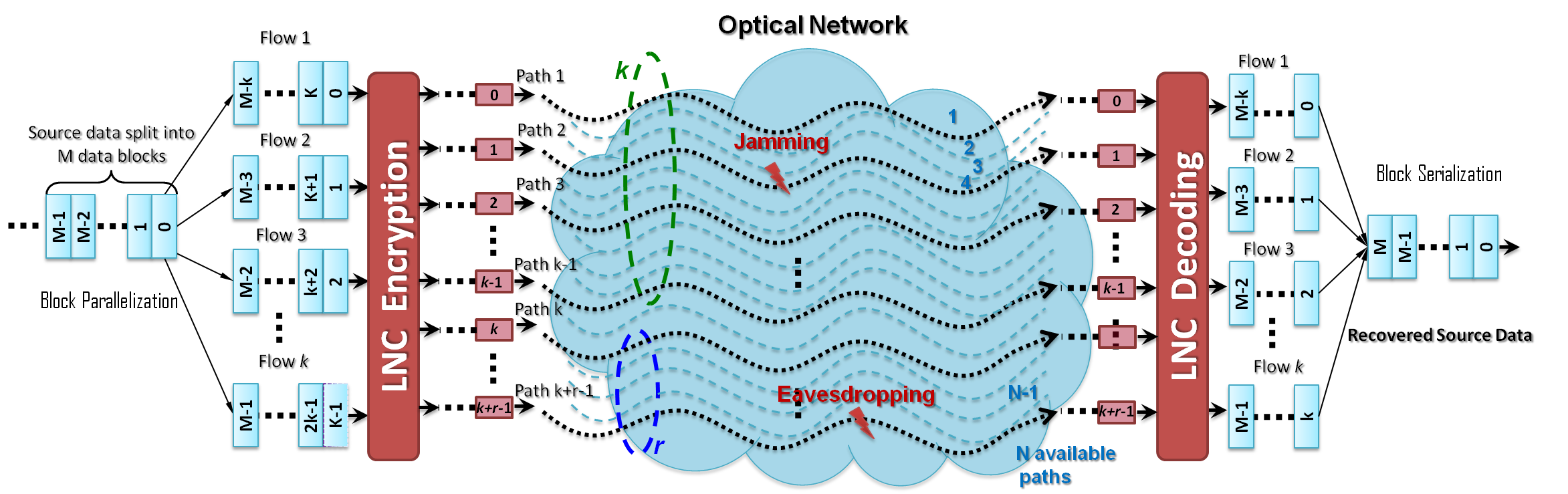}
\vspace{-0.0cm}
\caption{Reference network model with LNC in source and destination and security attacks in the network.}\label{networkM}
\vspace{-0.5cm}
\end{figure*}

\par In this paper, we study the effectiveness of LNC to balance the demands of reliable transmission and security in optical networks. To this end, we combine the coding process with parallelization of the source data stream, matching the generation size to the number of paths allocated in the optical network.  In our approach, a serial data stream from the source is parallelized into $n=k+r$ lightpaths, whereby $k$ represents number of data blocks of the original (secret) data and $r$ is a number of redundancies (random data sources) generated at the source to protect a secret data against wiretap security attacks. In addition, we study how the path selection in the optical layer impacts the security overall. Specifically, we study two path selection methods in circuit switched (WDM) transmission networks: 1) optimized paths, and 2) arbitrarily available paths. The results show that LNC and a random path selection method can effectively balance and improve both the security and reliability of optical transmission.   

\section{Modeling and analysis}
\subsection{Reference network model}

\par Fig.~\ref{networkM} shows our reference network model. The data traffic is modeled as a binary sequence, whereby original data can be decomposed into \emph{M data blocks} of the same length. Every data block is then parallelized and distributed in a round robin fashion to $k$ sub-flows at the source nodes. This is akin to the high speed Ethernet systems with Multiple Lane Distribution (MLD) \cite{802.3ba}. For instance, in an MLD Ethernet network an incoming serial flow consists of multiple 66b data blocks distributed over multiple lanes, where, for instance, $k=4$ lanes of $10Gbps$ each can be utilized to reach the total transmission rate of $40Gbps$ \cite{802.3ba}. 

Without loss of generality, and for simplicity, Fig.~\ref{networkM} illustrates how LNC is performed at the end-system, i.e., at source and destination, and the assumption is that there is no coding in the middle of the network\footnote{This is a valid assumption here, since additional coding inside the network can only increase the security and reliability of the data transmitted \cite{Pfennig:2013, Wang:2013}.}. The linear coding process is performed over a field $F_{2^m}$, where $2^m$ is the field size. Here, each data block is decomposed into \emph{symbols}, whereby each symbol has the same length of $m$ bits. Symbols encoded with the same set of coding coefficients are referred to as a \emph{generation}, while the number of original information symbols $k$ encoded with the same set of coding coefficients defines the \emph{generation size}. Here, new coding coefficients are utilized for each new set of data blocks from $k$ parallel sub-flows. Thus, the source data of $M$ data blocks result in $\frac{M}{k}$ generations.

\par In this network system, we assume that the number of outgoing encoded flows $n$, corresponding to the number of required optical paths $\xi$ over optical network, i.e., $n=\xi$. Thus, the source sends the encoded data over $\xi$ out of all $N$ available paths in the network. Generally, the number of outgoing encoded flows $n$ can be equal to or larger than number of incoming flows $k$, $n\geq k$, which is of special importance here. We refer to $r=n-k$ symbols sent as \emph{redundancy}.  Thus, a reliable transmission of one generation is possible since only $k$ out of $n=k+r$ arbitrary encoded symbols/data blocks from the same generation of any of $n$ paths are required for a successful decoding at the receiver. This provides protection against jamming attacks, when an attacker disrupt only $\nu\leq r$ data blocks from each generation. At the same time, a eavesdropping can be successfully defeated by LNC due to the fact that a wiretapper requires at least $k$ encoded data blocks from the same generation and knowledge about coding coefficients to recover a secret data.

\par A final feature in our reference network is the path selection. We assume that a network provides $\mathcal N$ paths between source and destination, while at most $N$ optical paths are available (i.e., not blocked) for a transmission request. For a successful data transfer, the choice of $n$, $n \leq N$ paths in the network can be either optimal (P-OPT) or arbitrary (P-RND). An optimization goal maybe to find $n$ shortest paths, with a minimal differential delay, or similar. To this end, offline and online tools maybe necessary to calculate paths, such as IETF Path Computation Element (PCE). On the other hand, any available path can be chosen for simplicity, or in absence of optimization tools. We study both methods and evaluate the benefits of path selection for security and reliability purposes, since depending on the location of the attacked edge, different path selection strategies will yield different performance. 

\subsection{Model Preliminaries}
\par The network is represented as a directed and acyclic graph $G (V, E)=G$, where  $V$ and $E$ are vertex set and edge set, respectively. The source and destination nodes are denoted as $s\in V$ and $d \in V$, respectively. A distinction is made between incoming and outgoing edges of an arbitrary node $v\in V$, which are denoted as a set $\mathcal E_{in} (v)$ and  $\mathcal E_{out}(v)$, respectively. In our model, $|\mathcal E_{in} (v)|\ =\sum_{i=1}^{u_{in}}u_{in_i}\cdot c_{in_i}$, if the optical switch supports $u_{in}$ input interfaces and each input interface $i$ provides $c_{in_i}$ different available wavelengths. Similarly, we define the outgoing links as $|\mathcal E_{out} (v)|=\sum_{j=1}^{u_{out}}u_{out_j}\cdot c_{out_j}$. An incoming link of a node $v$ is denoted as $head (e_i)= v$, where $1\leq i\leq |\mathcal E_{in} (v)|$. An outgoing link $e_j \in E$ of a node $v$ is denoted as $tail(e_j)=v$, where $1\leq j\leq |\mathcal E_{out} (v)|$. Since each edge and each link on a wavelength $\lambda_x$ in the network $G$ connects two nodes, e.g. $v'$ and $v''$, we denote each edge as $e_{v'v''}$ and each link as $e_{v'v''}(\lambda_x)\in e_{v'v''}$, respectively. The refined definition of link is $tail(e_{v'v'',j}(\lambda_x))=v'$ and $head (e_{v'v'',i}(\lambda_x))= v''$, $1\leq x\leq c_{e_{v'v''}}$, where $c_{e_{v'v''}}$ is capacity of edge measured in number of available wavelength for transmission. In other words, each edge $e_{v'v''}$ provides $c_{e_{v'v''}}$ parallel links between nodes $v'$ and $v''$. 

\par In the proposed system, the \emph{time unit (tu)} can be modeled as a discrete time based on the link capacity of the physical link and analyzed as transmission delay of one data block. Let us assume that at time $t$, the incoming symbols at source node $s\in V$ are generated by the processes $X^t(s, i)$ on every edge $i$ considered as an incoming link of a node $s$ and denoted as $e_i$, $head (e_i)= s$ and $1\le i\le k$. The incoming symbols are encoded into symbols $y(e_j)$, $1\le j\le n$, and sent out on each outgoing links $e_j \in E$ of a node $s$, $tail(e_j)=s$. In this model, the LNC encoder buffers incoming symbols from all incoming flows in parallel, and encodes the same with simple linear coding by using coding coefficients $a_{i, e_j}$ from the finite field $F_{2^m}$. Thus, the signal carried on an outgoing link $e_j$ of source $s$ at time $t+1$ is:
    \begin{equation}\label{source}
    \forall e_j: tail(e_j)=s:   \quad Y^{t+1}(e_j) = \sum_{i=1}^{k} a_{i, e_j} \cdot X^{t}(s, i)
\end{equation}

The decoder at receiver starts decoding as soon as one generation is complete, i.e., $k$ data blocks from the same generation are received. The decoder is able to decode one generation at a time by running Gaussian elimination. After decoding, the data blocks are serialized and leave the decoding functional block in the correct order. The decoded information at time $t+t_{\delta}$ on edge $i$ in LNC decoder at the destination $d$ is modeled as \cite{Koetter:TON:2003}:
\begin{equation}\label{decodingDest}
\forall e_i: tail(e_i)=d: \quad Z^{t+t_{\delta}}(d, i) = \sum_{j=1}^{n} b_{i, e_j } \cdot Y^{t}(e_j)
\end{equation}
, where $t_{\delta}$ is decoding interval and $b_{i, e_j}$ is decoding coefficients from the finite field $F_{2^m}$.

\subsection {Optical Path Selection}
\par We assume that a network provides at most $\mathcal N$ optical paths between source $s$ and destination $d$, whereby these paths are collected in path set $\Psi$. At the moment of transmission request, only $N$ out of $\mathcal N$, $\mathcal N\geq N$, paths are available. All $\mathcal N$ existing optical paths in $G$ are arranged in the vector $\vv{P}=\begin{pmatrix}
 P_0 \
P_1 \
 ... \
P_{\mathcal N-2} \
 P_{\mathcal N-1} \
\end{pmatrix}$ and vector $\vv{d}=\begin{pmatrix}
d_0 \
d_1 \
 ... \
d_{\mathcal N-2} \
d_{\mathcal N-1} \
\end{pmatrix}$
with related probabilities, that any path $\mathcal P_l(s,d)$, $l = 0, ..., \mathcal N-1$, between source and destination is available, and related end-to-end delays, whereby $\vv d$ is sorted in the ascending order. Here, $P_l$ and $d_l$ describes the same path $\mathcal P_l(s,d)$, $l = 0, ..., \mathcal N-1$. Generally, only $\xi=n$, $\xi \leq N$ and $\xi \leq|\mathcal E_{out}(s)|$ optical paths are required for transmission, out of $N\leq\mathcal N$ available paths. Given the paths available, every intermediate node $v$ can forward $m$ incoming data blocks from source over pre-configured  set of $|\mathcal E_{out}(v)|\geq m$ available outgoing links. This is feasible, when the min-cut between source $s$ and destination $d$ in $G$ is at least $\xi=n$, $min\_cut\{s,d\}\geq n$, so that the set of all available parallel paths $\Psi_N\subset\Psi$ in $G$ is larger than or equal to the number of encoded flows, $n$, at the source $s$, i.e., $N\geq n$. 

\par As previously mentioned, paths can be chosen from the set of all available paths in either arbitrary fashion, or, as it is more commonly the case, according to an optimization objective, for instance to minimize the differential delay.  We refer to these two methods as P-RND and P-OPT, respectively. To calculate the probabilities that a certain path, and a set of paths, is chosen for transmission, let us now assume that there is a collection, or a super-set $\mathcal A$, which contains $\mathfrak a= C(|\Phi|,\gamma):=\binom{|\Phi|}{\gamma}$ path sets $\mathcal A_{\alpha}$, $1\leq\alpha\leq\mathfrak a$, while $\Phi$ can be a collection of all existing paths $\Psi$ or a collection of available paths $\Psi_N$, i.e., $\Phi:=\Psi$, $|\Phi|=\mathcal N$ or $ \Phi:=\Psi_N$, $|\Phi|=N$, and $\gamma$ can be a number of available paths $N$ or the number of required for transmission paths $\xi$, i.e., $\gamma:=N$ or $\gamma:=\xi$. Thus, each set $\mathcal A_{\alpha}$ consists of $\gamma$ elements. In contrast, set $B_{\alpha}=\Phi\backslash A_{\alpha}=\{\mathcal P_{l} (s, d)| (\mathcal P_{l} (s, d)\in B_{\alpha}) \land (\mathcal P_{l} (s, d)\notin A_{\alpha}) \}$ from collection $\mathcal B$ is the $\alpha^{th}$ set of remaining $|\Phi|-|A_{\alpha}|$ elements, which are not in the $\alpha^{th}$ combination $A_{\alpha}$. Thus, the probability $P''(\alpha,\gamma, \Phi)$, that $\gamma $ out of $|\Phi|$ paths from set $A_{\alpha}$ and not  from set $B_{\alpha}$ are selected, is defined as 
\begin{equation}\label{PrPathComb}
P''(\alpha, \gamma, \Phi)=\prod_{i=1,\atop \mathcal P_{l,i}\in A_{\alpha}}^{\gamma}P_{l,i}(\alpha)\prod_{t=1,\atop \mathcal P_{l,t}\in B_{\alpha}}^{\mathcal |\Phi|-\gamma}(1-P_{l,t}(\alpha))
\end{equation}
, whereby the probabilities $P_{l,i}(\alpha)$ and $P_{l,t}(\alpha)$ are probabilities of a path $\mathcal P_{l}(s,d)$, $l=1,2,...,\mathcal N$, which are collected in $A_{\alpha}$ and $B_{\alpha}$, respectively, to be available, while indexes $i$ and $t$ show the sequence number of a path $\mathcal P_{l} (s, d)$ in the $\alpha^{th}$ paths combination from $\mathcal A$ and $\mathcal B$, i.e., $A_{\alpha}$ and $B_{\alpha}$, respectively.
\par As a result, the network provides $N=j$ parallel available paths with probability $\hat P(N=j,\Phi)$ defined as follows
\begin{equation}\label{PrNPath}
\hat P(N=j,\Phi)=\sum_{\alpha=1}^{\binom{|\Phi|}{j}}P''(\alpha, j, \Phi)
\end{equation}
, where the $\alpha^{th}$ set from the collection $\mathcal A$ contains all $\binom{|\Phi|}{j}$ path combinations of $N=j$ available paths with related probabilities described by vector $\vv P$.   

When there are $N<\xi$ available paths, the transmission is not possible and the transmission request will be blocked with probability $P_B(\xi)$, i.e.,
\begin{equation}\label{RequestBlocking}
P_B(\xi, \Phi)=\sum_{j=0}^{\xi-1}\hat P(N=j, \Phi)
\end{equation}
\par To find a set of optimal paths (P-OPT) with respect to minimum differential delay, we arrange all $N$ available paths from each out of $\mathfrak a=\binom{|\Psi|}{N }$  combination $ A_{\alpha}$, $1\leq\alpha\leq\mathfrak a $ in a delay vector  $\vv{d_{\alpha}}=\begin{pmatrix}
d_0 \
d_1 \
 ... \
d_{N-2} \
d_{N-1} \
\end{pmatrix}$, sorted in the ascending order of their corresponding path delays, i.e. $d_{l'}\leq d_{l'+1}$, of each available path $\mathcal P_{l'}(s,d)$, $0\leq l' < N-1$. Thus, the optimal paths are the paths with delays $d_0, d_1,...,d_{\xi-1}$.

\par When paths are chosen randomly, i.e., P-RND, only one out of $\mathfrak a =C(N,\xi):=\binom{N}{\xi}$ path combinations from $\mathcal A^R$ is selected for transmission; for example, paths from set $\mathcal A^R_{\alpha}\in \mathcal A^R$, $1\leq\alpha\leq C(N,\xi)$. Since we assume that all $C(N,\xi)$ path combinations have the same probability to be selected for transmission, the probability of an arbitrary paths combination is defined as $\frac{1}{C(N,\xi)}$.

\subsection{Security Analysis}
\subsubsection{Treat model}
\par Let us assume that there are two collections, or super-sets of edges,  $\mathbf W^E$ and $\mathbf W^J$ of sets $\mathfrak W^E$ and $\mathfrak W^J$ of wiretap edges in $G$, on which an attacker can perform eavesdropping or jamming attacks. Moreover, we assume that the wiretap edges are static, i.e., they were selected by a wiretapper only once and not changed in course of our analysis. The attacker can access only one $\mathit w^{th}$ set from collection $\mathbf W^E$ or $\mathbf W^J$, i.e., $\mathfrak W^E_{\mathit w}$ or $\mathfrak W^J_{\mathit w}$. However, we assume that both types of attack can be performed simultaneously, but on different edges, e.g., $e_{v'v''}\in\mathfrak W^E_{\mathit w}$ and $e_{v'v''}\notin\mathfrak W^J_{\mathit w}$. Thus, an attacker can attack at most $\hat{\mathbf w}=\mathbf w^E+\mathbf w^J$, $\mathbf w^E = | \mathfrak W^E_{\mathit w_E}| $ and $\mathbf w^J = | \mathfrak W^J_{\mathit w_J}| $, edges in $G$ at the same time. Moreover, we consider the worst case scenario and assume that an attacker is able to access all $c_{e_{v'v''}}$ wavelengths over one fiber link, when it has an access to edge $e_{v'v''}$ between nodes $v'$ and $v''$. As a result, the total number of wiretap links is $c_W= \sum_{\atop\forall e_{v'v''}\in\mathfrak W}^{}{c_{e_{v'v''}}}$, where $\mathfrak W=\mathfrak W^E_{\mathit w}\bigcup\mathfrak W^J_{\mathit w}$.

\subsubsection{Amount of data blocks attacked}
\par When at most $\mathbf w$ edges $e_{v'v''}\in E$, with capacity $c_{e_{v'v''}}$ each, are wiretapped, i.e., $\forall e_{v'v''}\in\mathfrak W^E_{\mathit w}$ or $\forall e_{v'v''}\in\mathfrak W^J_{\mathit w}$, the number of wiretap  paths $y^w$ is defined as a number of wiretap links utilized for transmission in a certain path combination. On the other hand, the probability that the wiretap paths are utilized for transmission depends on path probability described by vector $\vv P$, i.e., the probability that a wiretap paths are available. Moreover, the expected number of utilized wiretap paths $\bar{\mathcal Y}$ is a function of an occurrence of a certain set of available paths, which has a certain size $N$, and the final path selection for data transmission, whereby some of them can be a wiretap paths. We define the number of wiretap paths $y^{w}$ in a known, $\alpha^{th}$,  path combination utilized for transmission as 
\begin{equation}\label{numWpaths}
y^{w}(\alpha, \xi)=\sum_{l'=0, \atop \mathcal P_{l'}(s,d)\in\mathcal A_{\alpha}}^{\xi-1} Q(\mathcal P_{l'}(s,d))
\end{equation}
, where $\xi$ is the number of paths in each path combination $\mathcal A_{\alpha}$ utilized for transmission. Since a wiretapper can gain only one informative data block from each path regardless of the number of wiretap links belonging to this path, each path $\mathcal P_{l'} (s, d)$ can be considered only once as wiretap path in selected set $\mathcal A_{\alpha}$. To constraint that, we define a function $Q(\mathcal P_{l'}(s,d))$ as follow
 \begin{equation}\label{getQ}
\scalebox{0.9}{\begin{minipage}{2\columnwidth}
$Q(\mathcal P_{l'}(s,d))=\begin{cases}
1,& \text{if} \sum\limits_{\forall e_{v'v''}\in\mathfrak W}\sum\limits_{x=1}^{c_{e_{v'v''}}}z(e_{v'v''}(\lambda x),\mathcal P_{l'}(s,d))>0\\
0, &\text{else }
\end{cases}$
\end{minipage}}
\end{equation}
, where $\mathfrak W$ is a set of wiretap edges in network, i.e. $\mathfrak W=\mathfrak W^E_{\mathit w}$ or $\mathfrak W=\mathfrak W^J_{\mathit w}$ and consists of $\mathbf w$ edges , i.e., $\mathbf w:=\mathbf w^E$, $\mathbf w^E = | \mathfrak W^E_{\mathit w}| $ or $\mathbf w:=\mathbf w^J$, $\mathbf w^J = | \mathfrak W^J_{\mathit w}| $. The function  $z(e_{v'v''}(\lambda x), \mathcal P_{l'}(s,d))$ used in Eq.~\eqref{numWpaths} is a logical function, which shows whether a wiretap link $e_{v'v''}(\lambda x)$ is on the transmission path $\mathcal P_{l'}(s,d)$, i.e.,
\begin{equation}\label{getZ}
z(e_{v'v''}(\lambda x),\mathcal P_{l'} (s, d))=\begin{cases}
1,  & \text{if } e_{v'v''}(\lambda x)\in\mathcal P_{l'} (s, d)\\
0, &\text{else }
\end{cases}
\end{equation}

\par In case of optimal path selection, i.e., P-OPT, the wiretap path $\mathcal P_{l'}(s,d)$, $0\leq l'<\xi-1$, is utilized, when it is available, i.e., $\mathcal P_{l'}(s,d)\in \Psi_N$, and sorted in delay vector $\vv d_{\alpha}$ as $d_{l'}\in\{d_0, d_1,...,d_{\xi-1}\}$. Thus, $\forall e_{v'v'',i}: e_{v'v'',i}\in\mathfrak W_{\mathit w}$, the number of utilized wiretap paths is defined as follows
\begin{equation}\label{OPT}
\bar{\mathcal Y}_{opt}=\sum_{j=\xi}^{\mathcal N}\sum_{\alpha=1}^{C(\mathcal N, j)}\tfrac{P''(\alpha, j, \Psi)}{1-P_B(\xi,\Psi)}\cdot y^w(\alpha, \xi)
\end{equation}
, where  $C(\mathcal N, j)=\binom{\mathcal N}{j}$ defines the number of all possible combinations of $N=j$ available paths, while the probability of $\alpha^{th}$ combination, which consists of $N=j$ available paths, $P''(\alpha, j, \Psi)$, can be determined by Eq.~\eqref{PrPathComb}. Since transmission is only successful, if there are enough available paths, i.e., $\xi\leq N$, we only consider combinations of $j\geq\xi$ available paths, which occur with probability $(1-P_B(\xi,\Psi))$, see  Eq.~\eqref{RequestBlocking}.

\par 
In case of random path selection, i.e., P-RND, $\xi$ arbitrary available paths can be utilized for transmission with equal probability, whereby any $\xi$ paths out of $\mathcal N$ existing paths can be selected with equal probability, if these $\xi$ paths are available at the moment of transmission request with probability $\hat P(N=\xi, \Psi)$. Thus, the expected number of utilized wiretap paths $\bar{\mathcal Y}_{rnd}$ is 
\begin{equation}\label{RND}
\bar{\mathcal Y}_{rnd}=\sum_{\beta=1}^{C(\mathcal N,\xi)}\tfrac{P''(\beta, \xi, \Psi)}{\hat P(N=\xi, \Psi)}y^w(\beta, \xi)
\end{equation}
, where $y^w(\beta, \xi)$, $P''(\beta, \xi, \Psi)$ and $\hat P(N=\xi,\Psi)$ are determined by Eqs.~\eqref{numWpaths},~\eqref{PrPathComb} and~\eqref{PrNPath}, respectively.

To transmit all $M$ data blocks from the same secret data, the source utilizes $\xi=n$ parallel reserved paths. Thus, $\Lambda(y)$ data blocks can be collected or disrupted by a wiretapper from $y$ wiretap paths, while at most $M/\xi$ data blocks are transmitted over each wiretap link $e_{v'v'',i}$, i.e. $e_{v'v'',i}\in\mathfrak W^E_{\mathit w}$ or $e_{v'v'',i}\in\mathfrak W^J_{\mathit w}$ and $e_{v'v'',i}(\lambda_x)\in\mathcal P_{l'}(s, d)$. As a result, the total number of wiretap data blocks $\Lambda(y)$ is defined as  follows
\begin{equation}\label{seenGenWDM}
\Lambda(y)=\frac{M\cdot y}{\xi}
\end{equation}
, where the variable $y$ describes the number of wiretap paths utilized for transmission and, depending on path selection method and security attack, can be defined by Eqs.~\eqref{numWpaths}, \eqref{OPT} or \eqref{RND}, i.e., $y:=y^{w}(\alpha, \xi)$ or $y:=\bar{\mathcal Y}_{opt}$  or $y:=\bar{\mathcal Y}_{rnd}$. To protect against jamming attacks, we propose to send over $r$ additional paths $r=\Lambda (y)$ additional redundant data blocks generated at the source. To evaluate the ability of an attacker to recover or to disrupt a complete generation, we define the expected number of attacked data blocks from the same generations as 
\begin{equation}\label{seenGen}
\Lambda^*(y)=\tfrac{y}{k}
\end{equation}
, where $k$ is a number of data blocks required for successful decoding at destination.

\par Path selection methods has impact on security. To analyze this, let us consider the worst case scenario, where there is only one wiretap edge in the network, but most of the paths allocated to transmission go over it, thus leading to what we refer to as \emph{catastrophic security treat}. As a measure of catastrophic security treat, we define two probabilities related to eavesdropping and jamming attacks. For eavesdropping, we evaluate the probability that a wiretapper is able to receive over one wiretap edge $e_{v'v''}\in\mathfrak W^E_{\mathit w}$ at least $\nu=k$ LNC encoded data blocks from each generation and to encode the whole secret data. In case of jamming attacks, the wiretapper should be able to disrupt at least $\nu=r+1$ data blocks from each generation transmitted over edge $e_{v'v''}\in\mathfrak W^J_{\mathit w}$ and, thus, to prevent successful decoding at the destination. In other words, the secret data can experience a catastrophic security treat, when $\nu$ utilized parallel path are established over the same wiretap edge $e_{v'v''}$. The probability of catastrophic security treat $\Theta_{opt}(\nu, \xi)$ in case of P-OPT over $\xi$ parallel paths is defined as 
\begin{equation}\label{CSOPT}
\Theta_{opt}(\nu, \xi)=\sum_{j=\xi}^{\mathcal N}\sum_{\alpha=1}^{C(\mathcal N, j)} \tfrac{P''(\alpha, j, \Psi)}{1-P_B(\xi,\Psi)}\cdot\mathcal X(\alpha, \nu, \xi) 
\end{equation}
, where function $X(\alpha, \nu, \xi)$ described by Eq.~\eqref{validX} is a logical function, which defines a valid for analysis path combinations, i.e., the path sets utilized for transmission, where at least $\nu$ paths are wiretap paths. 
\begin{equation}\label{validX}
\mathcal X(\alpha, \nu, \xi)=\begin{cases}
1,  & \text{if } y^{w}(\alpha,  \xi)\geq \nu\\
0, &\text{else }
\end{cases}
\end{equation}
Finally, the probability of catastrophic security treat $\Theta_{rnd}(\nu, \xi)$ in case of P-RND transmission method over $\xi$ random paths is defined as 
\begin{equation}\label{CSRND}
\Theta_{rnd}(\nu, \xi)=\sum_{\beta=1}^{C(\mathcal N,\xi)}\tfrac{P''(\beta, \xi, \Psi)}{\hat P(N=\xi,\Psi)} \mathcal X(\beta, \nu, \xi)
\end{equation}

\section{Performance evaluation}
\par We now show the theoretical results and validate the same by simulations. The data blocks for both path selection methods, i.e., P-RND and P-OPT, have the same size of $80$ byte and the secret data was composed of $20$ data blocks. The LNC process was implemented over finite field $F_{2^8}$ with symbol size $8$ bits. The transmission rate of each optical link was defined as $10$ Gbps. The traffic load was set to 1 data block per \emph{time unit} per lane at the source node and the network is lossless. The normalized confidence intervals show 95\% confidence and small values, and are not shown for clarity.

\par
\begin{figure} [t]
  \centerline{\includegraphics[width=1\columnwidth]{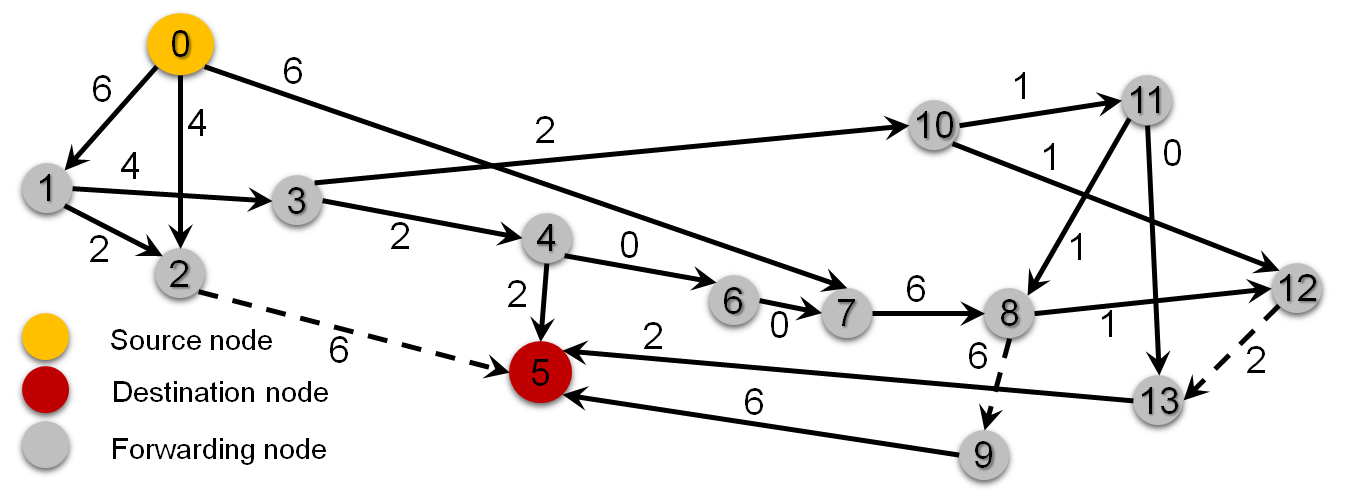}}
  \vspace{-0.2cm}
  \caption{Directed nsfNet with allocated edge capacities.}
  \label{topo}\vspace{-0.5cm}
\end{figure}

 For the simulations and analysis, we use the NSFnet (Fig.~\ref{topo}). Here, there are in total $\mathcal N=18$ parallel paths allocated to the source node $0$ and destination node $5$. The numbers on the edges in Fig.~\ref{topo} are used to denote the amount of parallel directed links allocated to the source-destination pair. All paths are sorted in the ascending order according to the path length, i.e., number of hops, and described by vector $\vv d=\{2,2,2,2, 3,3, 4,4,4,4,4,4,4,4, 5, 6, 7, 8\}$. Moreover, the parallel paths have different probabilities to be available at the moment of transmission request. We assume that these probabilities decrease with increasing path length, i.e., $\vv P=\{0.8,0.79,0.78,0.77,  0.70,0.69,  0.60,0.59, 0.58,0.57,0.56,\\0.55,0.54,0.53,  0.50,  0.45,  0.40,  0.35\}$. 
In case of P-RND, $\xi=n$ out of $N\leq\mathcal N$ available parallel paths are selected randomly. In case of P-OPT parallel transmission, the optical data flows sent in parallel over $\xi$ optimized (shortest) parallel paths. We assumed that an attacker has access to three edges $e_i\in\{ 2-5, 8-9, 12-13\}$ (dashed edges in Fig~\ref{topo}) and can perform different types of security attacks simultaneously on these edges, at different time points, and in different combinations. Based on this assumptions, we evaluate the amount of attacked data blocks for eavesdropping and jamming attacks. 

\par
\begin{figure} [t]
  \centerline{\includegraphics[width=1\columnwidth]{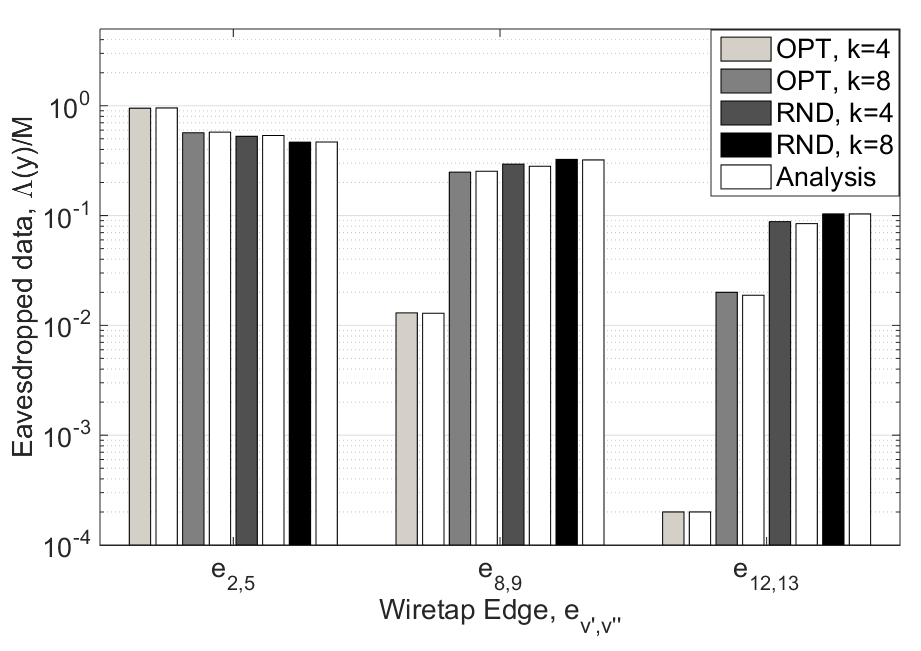}}
  \caption{Amount of wiretap data on a sample wiretap edge.}
  \label{edges}\vspace{-0.5cm}
\end{figure}

 Fig.~\ref{edges} shows the amount of eavesdropped data units per generation, $\Lambda(y)/M$, on a sample wiretap edge. Here, the transmission was implemented over $\xi=k$ parallel paths, i.e., without redundancy, while the generation size was set to $k=4$ and $k=8$ data units. In case of P-OPT with $4$ and $8$ parallel paths, the most data blocks, around $95\%$ and $57\%$, respectively, could be wiretap on the edge $e_{2,5}$, which is a high capacity edge (high number of parallel links) and belongs to the shortest paths between source and destination. In contrast, the amount of eavesdropped data blocks on other wiretap edges of longer paths is much smaller and increases with a number of parallel paths $\xi=k$ utilized for transmission. In case P-RND, the data blocks are distributed more uniformly over network and wiretap edges and impact of $k$ is not very significant, however the amount of eavesdropped data blocks slightly increases with increasing $k$ on the edges that belong to longer paths. The simulation results are very close to analytical results calculated with Eqs.~\eqref{OPT}, ~\eqref{RND} and ~\eqref{seenGenWDM}.

\begin{figure} [t]
  \centerline{\includegraphics[width=1\columnwidth]{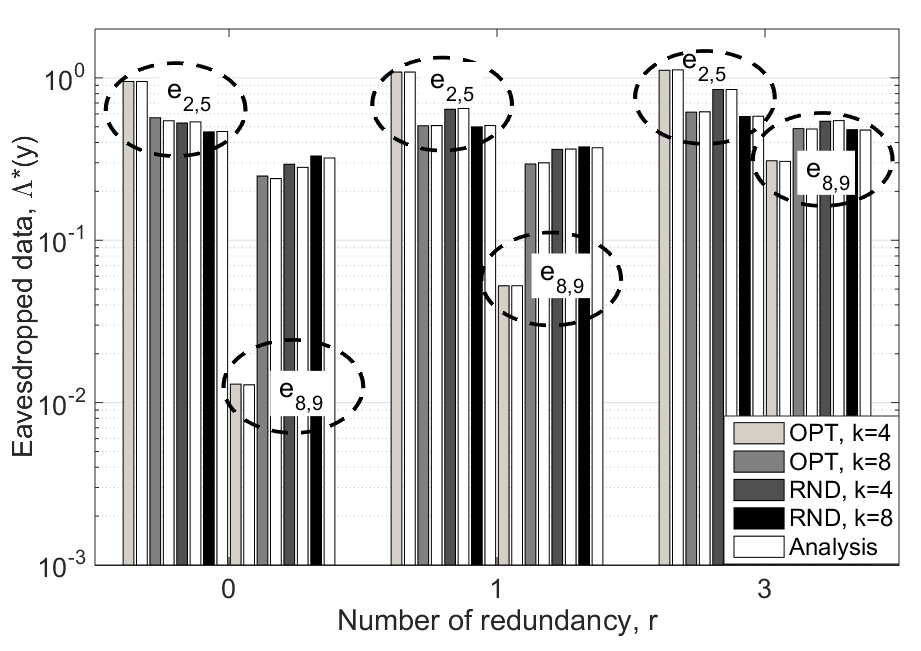}}
  \caption{Amount of wiretap data vs. coding redundancy.}
  \label{redund}\vspace{-0.5cm}
\end{figure}

\par Fig.~\ref{redund} shows a more detailed security analysis with respect to two specific wiretap edges, $e_{2,5}$ and $e_{8,9}$, which are chosen as they could deliver the most data blocks to a wiretapper (see  Fig.~\ref{edges}). Here, the amount of eavesdropped data blocks from the same generations $\Lambda^*(y)$ is presented as a function of a number of redundant data blocks, $r$, and generation size, $k$. Generally, the amount of eavesdropped data blocks increases with increasing $r$, and, in case of P-OPT with $k=4$, the wiretapper is able to receive the complete generation on edge $e_{2,5}$ and to successfully decode it, if at least one redundant data block is generated and sent over an additional path. In contrast, edge $e_{8,9}$ can provide at most $48.6\%$ of data blocks from the same generation with $r=3$. On the other hand, the P-RND disperses the secret data randomly so that the amount of wiretapped data blocks is lesser on edges belonging to the shorter paths, e.g., at most $84.8\%$ on the edge $e_{2,5}$ and transmission over $\xi=k+r=4+3$, and the amount of wiretap data on edges belonging to longer paths is larger, e.g.,  $55\%$ on the edge $e_{8,9}$ for the same $\xi$, compared to P-OPT. Also here, the simulation results match the analysis (Eqs.~\eqref{OPT},~\eqref{RND},~\eqref{seenGen}).

 \par
\begin{figure} [t]
  \centerline{\includegraphics[width=1\columnwidth]{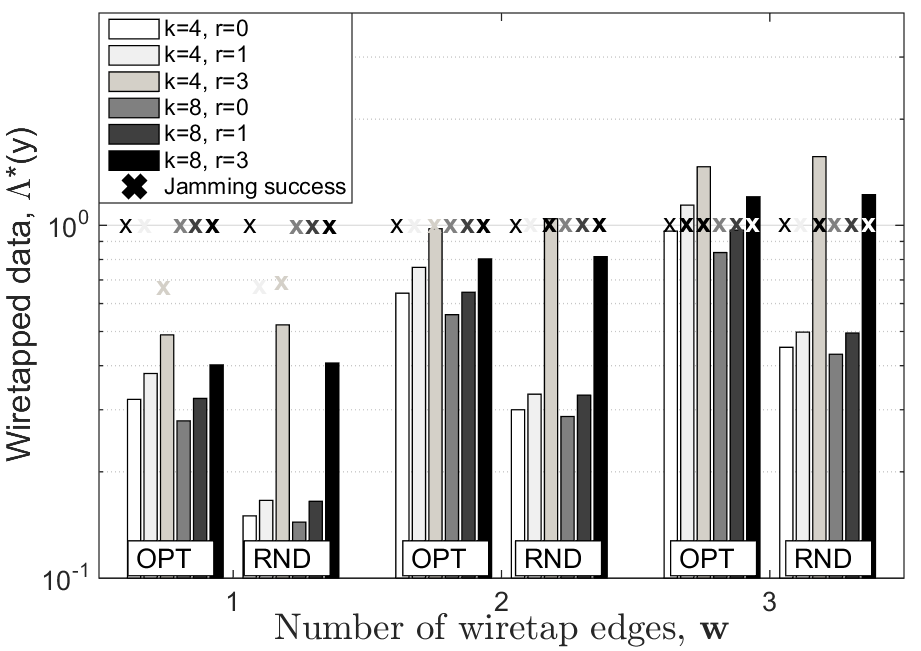}}
  \caption{Jamming success vs. wiretap data vs. wiretap edges.}
  \label{numEdge}
  \vspace{-0.3cm}
\end{figure}

 Next, we generate different combinations of wiretap edges, $e_i\in\{ 2-5, 8-9, 12-13\}$, where only one, two, or all three edges are attacked at the same time, i.e., $\mathbf w\in\{1,2,3\}$ and collection $\mathbf W$ consists of $3$, $3$ and $1$ elements, respectively. The results in Fig.~\ref{numEdge} present the mean number of eavesdropped data blocks from the same generation and jamming success, which were calculated over all possible combinations from $\mathbf W$. We show both the probability that a wiretapper is able to recover a generation and, as a result, a whole secret data, as well as the probability of jamming success. We assume the jamming was successful, if the attacker disrupted more than $r$ data blocks from the same generation and the receiver is not able to decode. The amount of eavesdropped data units in case of P-RND is much lower then in case of P-OPT for any $k$, when the number of redundant data blocks and, thus, number of paths is small, i.e., $0\leq r<3$. For a larger redundancy ($r=3$), P-OPT selects more shortest paths from a pool of non-wiretap paths, e.g., path $\{0-1-3-4-5\}$ in Fig.~\ref{topo}, and showed better performance. At the same time, P-RND distributed all data blocks nearly uniformly over all edges and thus utilized all wiretap edges. Here, the number of eavesdropped data blocks from the same generation $\Lambda^*(y)$ increases with increasing number of wiretap edges $\mathbf w$ attacked at the same time. The attacker can recover each generation by wiretapping of at least $2$ edges in the network, when $\xi=4+3$ paths are used. Overall, the amount of eavesdropped data units from the same generation and, thus, the probability that an attacker can recover the generation decreases with increasing generation size $k$. The jamming attacks were not successful, when the attacker had access to one edge only and $r=1$ or $r=3$ (Fig.~\ref{numEdge}). In this case, $66.4\%$ and $69.6\%$ of generations sent could not be decoded at destination, respectively, when P-RND was utilized and $k=4$.

\par
\begin{figure} [t]
  \centerline{\includegraphics[width=1\columnwidth]{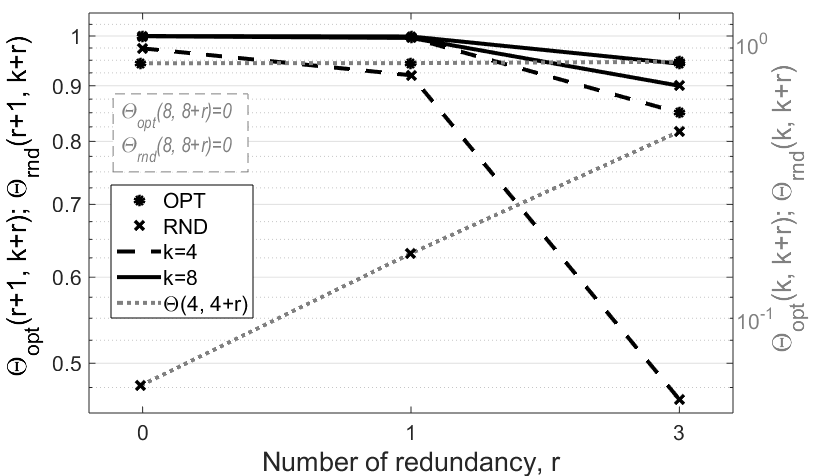}}
  \caption{Probability to eavesdrop a complete generation on $e_{2,5}$.}
  \label{E}
  \vspace{-0.3cm}
\end{figure}

 Finally, we evaluate the probability of catastrophic security treat, where we assume that the edge with a largest capacity allocated to the source destination pair, which is utilized to establish the shortest paths, is a wiretap edge, i.e., edge $e_{2,5}$. Fig.~\ref{E} shows the probability that an attacker can recover  or disrupt  a whole secret data by accessing edge $e_{2,5}$. The secret data recovery was only possible, when the generation size $k$ was small, i.e., $k=4$. For P-OPT, the probability for catastrophic security treat was nearly constant $~84\%$ for all levels of redundancy. For P-RND, the probability for catastrophic security treat increased from $~5\%$ to $~46\%$ with an increasing $r$. In contrast, the attacker gained no knowledge about secret data regardless of path selection method, when the generation size was set to $k=8$. On the other hand, an attacker could disrupt the complete secret data where no or only one redundant data block was generated to protect each generation. The increase in redundant data blocks decreases the probability for catastrophic security treat. P-RND with $k=4$ showed the best performance against jamming attacks, while the probability for catastrophic security treat decreased from $97\%$ with no redundancy to $46\%$ with $r=3$.

\section{Conclusion}
 In this paper, we studied the effectiveness of LNC to balance reliable transmission and security in optical networks. To this end, we combined the coding process with parallelization of the source data stream. We proposed and compared two WDM path selection methods: 1) optimized (shortest) paths, and 2) randomly selected paths. We analyzed and compared both the security capabilities against eavesdropping attacks, and reliability against jamming attacks. The results showed that LNC and parallel transmission over randomly selected paths significantly increases security against both eavesdropping and jamming compared with conventional WDM transmission, especially in the case, when a wiretapper has access to the shortest paths between source and destination. The combination of LNC and random path selection requires smaller number of redundancies to prevent jamming attacks. Finally, we showed that in absence of physical layer security measures, conventional WDM transmission can lead to a \emph{catastrophic security treat}, whereby an attacker can eavesdrop or to disrupt a whole secret data by accessing only one edge in a network.


\begin{thebibliography}{1}
\bibitem{Yuan:2014}
S.~Yuan and D.~Stewart, ``Protection of optical networks against interchannel
  eavesdropping and jamming attacks,'' in \emph{Computational Science and
  Computational Intelligence, International Conference on}, vol.~1, 2014, pp.
  34--38.

\bibitem{Furdek:2014}
M.~Furdek, N.~Skorin-Kapov, S.~Zsigmond, and L.~Wosinska, ``Vulnerabilities and
  security issues in optical networks,'' in \emph{Transparent Optical Networks,
  16th International Conference on}, 2014, pp. 1--4.

\bibitem{Cai:2011}
N.~Cai and T.~Chan, ``Theory of secure network coding,'' \emph{Proceedings of
  the IEEE}, vol.~99, no.~3, pp. 421--437, March 2011.

\bibitem{Fragouli:2006}
C.~Fragouli, J.-Y. Le~Boudec, and J.~Widmer, ``Network coding: An instant
  primer,'' \emph{SIGCOMM Comput. Commun. Rev.}, vol.~36, no.~1, pp. 63--68,
  Jan. 2006.

\bibitem{802.3ba}
``Iso/iec/ieee international standard for ethernet,'' \emph{ISO/IEC/IEEE
  8802-3:2014(E)}, pp. 1--3754, April 2014.

\bibitem{Pfennig:2013}
S.~Pfennig and E.~Franz, ``Secure network coding: Dependency of efficiency on
  network topology,'' in \emph{Communications (ICC), 2013 IEEE International
  Conference on}, June 2013, pp. 2100--2105.

\bibitem{Wang:2013}
J.~Wang, J.~Wang, K.~Lu, B.~Xiao, and N.~Gu, ``Modeling and optimal design of
  linear network coding for secure unicast with multiple streams,''
  \emph{Parallel and Distributed Systems, IEEE Transactions on}, vol.~24,
  no.~10, pp. 2025--2035, Oct 2013.

\bibitem{Koetter:TON:2003}
R.~Koetter and M.~Medard, ``{An Algebraic Approach to Network Coding},''
  \emph{IEEE/ACM Transactions on Networking}, 2003.

\end{thebibliography}
\end{document}